# Double electron resonance with two ensembles of nitrogen-vacancy centers in diamond

A. Chernyavskiy[1,2,3], I.S. Cojocaru[1,3,4], S.M. Drofa[1,2,3], P.G. Vilyuzhanina[1,2,3], A.M. Kozodaev[1,2,3],

V.G. Vins[6], A.N. Smolyaninov[4], S.Ya. Kilin[5,7], S.V. Bolshedvorskii[3,4], V.V. Soshenko[3,4] and

A.V. Akimov[1,3,4]

[1]Russian Quantum Center, Bolshoy Boulevard 30, building 1, Moscow, 143025, Russia

[2]Moscow Institute of Physics and Technology, 9 Institutskiy per., Dolgoprudny, Moscow Region, 141701, Russia

[3]P.N. Lebedev Institute RAS, Leninsky Prospekt 53, Moscow, 119991, Russia

[4]LLC Sensor Spin Technologies, 121205 Nobel St. 9, Moscow, Russia

[5]National Research Nuclear University "MEPhI", 31, Kashirskoe Highway, Moscow, 115409 Russia

[6]LLC Velman, 1/3 st. Zelenaya Gorka, Novosibirsk ,630060, Russia

[7]B.I. Stepanov Institute of Physics NASB, 68, Nezavisimosty Ave, Minsk, 220072 Belarus

email: a.akimov@rqc.ru

Nitrogen-vacancy (NV) centers in diamond are widely used in the development of a number of sensors. The sensitivity of these devices is limited by both the number of centers used and their coherent properties. While the effects on the coherent properties of paramagnetic impurities such as carbon 13-isotopes and p1 centers are rather well understood, the mutual interaction of NV centers, which becomes especially important in relatively dense NV ensembles, is less well understood. Here, we provide a systematic study of NV-NV interaction using a dynamical double electron-electron resonance sequence, making it possible to directly observe the interaction of NV centers. Two types of dynamical DEER sequences were considered, consisting of 3 and 4 pulses. The nature of the phase jump in the 3-pulse sequence was attributed to the effect of non-commuting rotations within the sequence. Both the phase of the state vector rotation and its amplitude decay were studied, thus presenting a complete picture of decoherence due to NV-NV interaction. It was shown that the rate of the state vector decay differed significantly from predictions for a spin ½ system. However, the decay rate observed in the DEER sequence remained a reliable indicator of the concentration of "bath spins" and could be used to measure NV center concentration, provided that the magnetic transition of NV centers is saturated.

## I.  INTRODUCTION

Double electron-electron resonance (DEER), also known as pulsed electron double resonance (PELDOR), is one of the pulsed electron paramagnetic resonance (EPR) methods used to probe the effect of the spin environment on the Hahn echo signal [1]. DEER was proposed in [2,3] and has become a widely used method for determining distances between paramagnetic centers in both glassy frozen solutions and crystals. Today, the 4-pulse sequence [4] is generally preferred over the 3-pulse DEER. DEER sequences with 5 and 7 pulses have been studied in [5]. In [6], the influence of pulse shape on interaction suppression was investigated, and an optimal pulse shape was identified. In systems with a fixed distance between spins, the Fourier transform is used to determine both the dipole and exchange interaction energies, as well as the distribution of distances between atoms [7]. When spins are randomly distributed within the sample, the DEER time trace exhibits exponential decay. The decay parameters can be used to determine the concentration of paramagnetic centers. The implementation of spin labels has enabled the application of this method to the study of protein molecules [8], as well as RNA [9] and DNA [10].

Nitrogen-vacancy (NV) centers in diamond [11] have several advantages for DEER studies. First, they allow optical polarization and spin state readout via fluorescence intensity, simplifying the experimental setup compared to traditional EPR systems. Second, NV centers exhibit high coherence time $T_2$ and longitudinal relaxation time $T_1$ even at room temperature. This enables longer experiments and the detection of weaker interactions. The DEER method was used in [12,13] to study the spectra of $p_1$ centers (also known as C-centers) in diamond and to determine the concentration of $p_1$ centers in [14,15]. In both cases, the interaction of "dark" spin $1/2$ impurities with NV centers was studied. Meanwhile, sensor applications often utilize relatively dense ensembles of NV centers, which are spin-1 systems and also exhibit spin-dependent fluorescence. In such dense ensembles, mutual interaction between NV centers cannot be neglected. Previously, such interactions in crystals were studied in [16], via forbidden transitions $|m_S = -1\rangle \rightarrow |m_S = +1\rangle$. The NV-NV DEER technique was also used to determine the distance between two closely spaced single NVs [17].

Here, we demonstrated the optically detected DEER signal detection of the selected orientation of "bright" NV spins by NV centers with different orientations. The previously observed distortion in the 3-pulse DEER data was attributed to a phase shift related to the non-commuting operations of microwave pulses in the DEER sequence. This distortion was described analytically and removed through the use of a 4-pulse sequence. The dynamics of the quantum state phase were studied and compared with existing theoretical expressions. Finally, DEER decay kinetics were studied for various numbers of excited spins by varying the excitation level of the environment. The developed approach can be extended as a method for measuring local NV concentrations, provided sufficient microwave power is available to flip all the environmental NV centers.

## II. MATERIAL AND EXPERIMENTAL METHODS

### A. Experimental setup

The experimental setup is depicted in Figure 1a. A laser diode beam with a wavelength of $\lambda = 532$ nm is focused on the diamond sample for fluorescence excitation and spin state readout. The diamond is attached to a parabolic concentrator to collect NV fluorescence. From the concentrator, the collected light passes through a filter, which blocks the laser radiation, and then reaches a photodiode. Two diamond samples with natural abundance of $^{13}C$ were used in the experiments: one with 1.36 ppm of NV centers, and another with 0.7 ppm. A portion of the laser radiation (10%) is deflected to a photodiode for measuring laser intensity. This signal may be utilized to reduce laser intensity noise in the final data.

NV center optical pumping is performed using a 520-nm laser diode. The laser diode controller operates in an on-off modulation regime mode to suppress excitation during MW pulses used for spin manipulation. A DC bias magnetic field created by a permanent magnet is applied to resolve magnetic transitions in the ODMR spectrum, associated with different orientations of NV centers in the crystal. The DC field is nearly aligned with the (111) axis of the diamond plate. The microwave signal path begins with the synthesis of an analytical signal in an FPGA, which is then sent to a dual-channel digital-to-analog converter. The resulting dual analog signal is upconverted to a carrier frequency using an ADF4351 integrated circuit board and then amplified by a 40 dB power amplifier Mini-Circuits ZHL-16W-43. Finally, the generated microwave pulses are delivered to the NV centers via coils arranged in a quasi-Helmholtz configuration [18] to excite magnetic transitions.

## B. Pulse sequences

Before proceeding with the DEER experiment, several preparation steps are necessary. The first step is tuning the bias magnetic field to resolve all magnetic resonances corresponding to different NV orientations in the optically detected magnetic resonance (ODMR) spectrum. For this purpose, the pulsed ODMR method is utilized. The first NV center is initialized in the $|m_S = 0\rangle$ state by a laser pulse. Then, an MW pulse is applied to transfer part of the population to one of the states $|m_S = \pm 1\rangle$. After that, the laser is switched on again, and the fluorescence intensity is read (see Figure 1b). If the MW field is resonant with one of the magnetic transitions, population transfer occurs and fluorescence decreases, because the $|m_S = \pm 1\rangle$ states are darker than the $|m_S = 0\rangle$ state. By sweeping the frequency of the MW pulse, the ODMR spectrum is obtained (see Figure 1c). In the spectrum, different triplets correspond to ensembles of NV centers with different orientations. Each triplet consists of three magnetic transitions, corresponding to the three possible projections of the nitrogen nuclear spin. The orientation of the NV center is selected by choosing the appropriate MW pulse frequency. If two different MW frequencies are used in a pulse sequence, two groups of NV centers can be addressed simultaneously, as is known from magnetometry experiments [19]. In this work, we refer to the NV spins used as detectors, on which the spin echo sequence is performed, as "spins A." "Spins B" correspond to NV centers with a different orientation, which produce a magnetic field affecting spins A during the DEER sequence. To generate this magnetic field effect, spins B are subjected to a $\pi$ pulse to flip their spin state during the DEER sequence.

NV centers with the largest Zeeman splitting are those aligned most closely along the magnetic field. These are selected as spins "A" because they exhibit longer coherence times in the echo sequence. In subsequent experiments, we chose to excite a single hyperfine transition for both spins A and spins B. For spins A, the rightmost resonance in the spectrum was selected, corresponding to the $|m_I = 1\rangle$ nitrogen nuclear spin substate. For spins B, the middle hyperfine transition was selected, corresponding to the nuclear spin projection $|m_I = 0\rangle$. This choice was made to improve pulse excitation efficiency and thus enhance the response of spins B in the DEER signal. However, in the case of high-power MW pulses, nuclear spin selectivity can be lost, as is observed in some of the following cases. It should also be noted that, since the applied magnetic field is relatively small and the corresponding Zeeman splitting $\gamma B$ $\gamma_e B = (2.8\,\text{MHz/G}) \cdot 23\,\text{G} \approx 65\,\text{MHz}$ is much smaller than the zero field splitting ($D = 2.7\,\text{GHz}$), the NV center's quantization axis remains nearly aligned with its symmetry axis.

In order to accurately define the resonant frequencies of spin A and spin B, the ODMR signal was clipped near the region of interest and fitted using a Lorentzian shape [14]:

$$L(f) = L_0 - \sum_{i=1}^{3} \frac{\gamma A_i}{(f - f_i)^2 + \gamma^2}. \tag{1}$$

Here $f_i$ is the central frequency of each peak, $A_i$ is the amplitude of each peak, $\gamma$ is half-width at half-maximum, and $L_0$ is the baseline signal.

After defining the pulse frequencies, it is necessary to determine the correct pulse durations to perform the DEER experiment. For this purpose, Rabi oscillations were measured for both spins A and B to calibrate the $\pi$-pulse durations corresponding to a selected microwave power level. For spins A, the pulse durations were set to: $t_\pi = 0.88$ μs, $t_{\pi/2} = 0.4$ μs with 180 mW of MW power used. The difference in times is due to delays in the electronics.

To estimate the average magnetic field strength induced by spins B on spins A during the DEER experiment, it is necessary to determine the fraction of spins B that are flipped by the corresponding $\pi$-pulse—that is, the pulse efficiency. This can be measured using a hole-burning experiment [20]. The hole at the level $|m_S = 0\rangle$ can be burned by applying the microwave $\pi$-pulse at $|m_S = 0\rangle \rightarrow |m_S = -1\rangle$ transition. Then, to define the leftover $|m_S = 0\rangle$ population after hole-burning $\pi$-pulse, the ODMR spectrum is obtained for the $|m_S = 0\rangle \rightarrow |m_S = +1\rangle$ transition range (see Figure 1d). Rabi frequency for probing is reduced to avoid power broadening. The amplitudes of ODMR resonances after hole-burning will be lower than in ODMR. Thus, by fitting ODMR, taking the difference in resonance amplitudes, pulse flip efficiency can be estimated using the following formula:

$$p_B = \frac{1}{4}\left(1 - \frac{\sum_{i=1}^{3} A_i}{\sum_{i=1}^{3} B_i}\right). \tag{2}$$

Here $A_i$ – the amplitude of the $i$-th peak obtained from the hole burning spectrum fit, $B_i$ – the amplitude of the $i$-th peak obtained from the ODMR spectrum fit, ¼ corresponds to the fact that population distributed equally between four crystallographic orientations. The result of such experiment (see Figure 1d) with 0.17 μs pump pulse duration is presented, here 68% of spins B or 17% of whole NV population was excited according to (2).

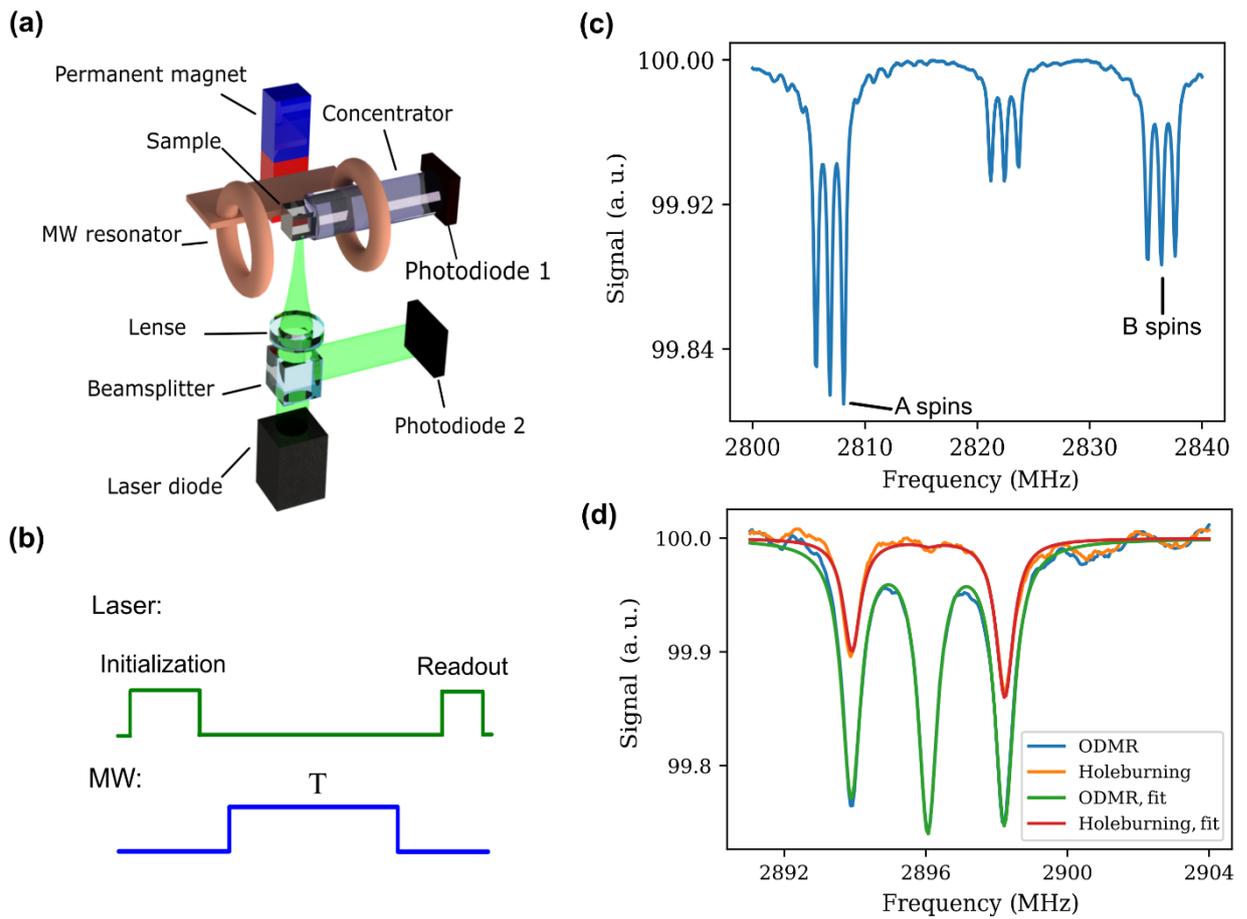

*Figure 1. a) Experimental setup, b) Scheme of the ODMR experiment, c) ODMR spectrum with the presence of both A and B spin transitions, d) ODMR spectra before (blue) and after (orange) hole burning; green and red plots are for Lorentzian fits. The difference in baselines was determined from the fit and subtracted.*

Since the echo sequence (Figure 2a) on spins A is a component of the DEER experiment (Figure 2b), its duration must be carefully selected to maximize the signal originating from spins B. In the case of low nitrogen concentration, as in the samples used in this study, the dominant source of decoherence in the echo signal is the $^{13}$C nuclear spin bath [22], which results in a stretched exponential (power-exponential) decay envelope modulated by periodic revivals (Figure 2c). At the revival maxima, the total echo duration corresponds to an even multiple of the $^{13}$C Larmor precession period, during which the decoherence contribution from $^{13}$C spins is partially suppressed. The first revival occurring at $\tau = 41\,\mu s$ was used for the DEER experiment due to its longer integration time window and the higher amplitude of the echo signal at that point.

To perform accurate coherence decay measurements in echo and DEER experiments, full state tomography must be carried out at the readout stage. This is achieved by varying the phase of the final $\pi/2$ microwave pulse relative to the first applied microwave pulse. The pulse phase notation used "x", "y","-x","-y" corresponds to pulse phases altered by 0, 90, 180, and 270 degrees, respectively. On the Bloch sphere, an $\pi_x/2$ pulse rotates the state vector around the x-axis, while a $\pi_y/2$ pulse rotates around the $y$-axis [21].

In Figure 2c, echo time traces are presented for two projections of the state vector on the Bloch sphere, namely $x$ and $y$, obtained from a pair of echo signals with opposite polarity.

$$E_x = I_y - I_{-y},$$
$$E_y = I_x - I_{-x}. \qquad (3)$$

where $I_\alpha$ is the echo signal from an experiment with the final rotation in the sequence is performed around the axis $\alpha$. The value proportional to the length of the state vector then can then be defined as:

$$E = \sqrt{E_x^2 + E_y^2}. \qquad (4)$$

As shown in Figure 2c, the $x$-projection is not zero, indicating that the spin state not only decays but also slowly rotates around the $z$ axis. This may be caused by a detuning of the microwave frequency from exact resonance, which can occur after the laser is switched off and the diamond temperature slightly changes. In the presence of detuning, the $\delta\nu$ state will rotate around the $z$ axis by the angle $\varphi$:

$$\varphi = 4\pi\tau\delta\nu \qquad (5)$$

### III. RESULTS AND DISCUSSION

#### A. DEER spectrum

To obtain the DEER spectrum, the echo sequence for spins A is modified such that a $\pi$-pulse is applied simultaneously to both spins A and spins B (Figure 2b), forming the so-called 3-pulse DEER sequence. In the case of a single spin A and a single spin B, both initially polarized at the start of the DEER experiment, the $\pi$-pulse applied to spin B inverts its state, thereby altering the energy splitting of spin A via dipole-dipole coupling. As a result, spin A acquires an additional dynamic phase without undergoing decoherence. However, for spin ensembles randomly distributed throughout the sample, the overall ensemble state becomes a statistical sum over many dipole coupling realizations, leading to decoherence of the spin A state vector. In the experiment described below, the $|m_S = 0\rangle \rightarrow |m_S = -1\rangle$ transition was used for both ensembles. At the start of each measurement, both ensembles were initialized in the $|m_S = 0\rangle$ state before being manipulated by microwave pulses.

The application of a flipping pulse on spins B affects the coherence of spins A, resulting in a change in the fluorescence signal from spins A. At the same time, the inversion pulse on spins B also modifies their own fluorescence intensity. However, it is not possible to distinguish between the fluorescence contributions from the two spin groups. Full state tomography of spins A enables extraction of the DEER signal. The fluorescence contribution from spins B remains the same across all measurements. However, the revival of the echo signal from spins A exhibits opposite signs depending on the phase of the final pulse: signals acquired with $x$, $-x$ and $y$, $-y$ phases, respectively, show different behaviors. Specifically:

$$D_x = I_y - I_{-y} \tag{6}$$

$$D_y = I_x - I_{-x} \tag{7}$$

$$D = \sqrt{D_x^2 + D_y^2} \tag{8}$$

$$\varphi = arctan\left(\frac{D_y}{D_x}\right) \tag{9}$$

here $I_\alpha$ is the obtained echo signal while the phase of the last MW pulse is $\alpha$.

Since for all four experiments, the spin B fluorescence is the same, in formulas (7) and (8), it is subtracted from the final signal. Thus $D_x$ and $D_y$ do not contain spin B fluorescence. The $D$ value from formula (8) will reveal coherence decay and will not contain the effect of phase shift, which can be resolved separately using formula (10).

In addition to this, three types of experiments were performed for each phase: in the first, MWs were applied only to spins B, in the second, MWs were applied only to spins A, and in the third, both spins were excited with MWs. In the first case, after the substitution of two phases, a zero signal occurs, because it is the difference between two ODMR spectra. In the second case, a constant echo signal occurs, in the third – a DEER signal. In Figure 2d, the results of all three measurements are shown and divided by the average echo signal 0.023 V. The NV triplet appears only in the third plot of DEER, which means that the fluorescence of spins B is subtracted correctly.

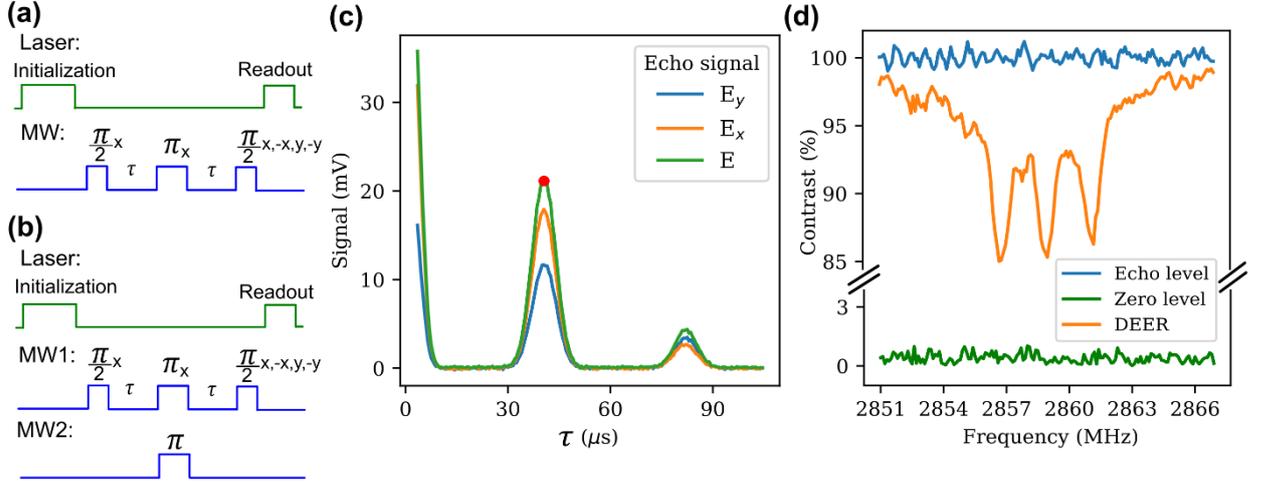

*Figure 2. a) Scheme of the spin echo experiment, b) Scheme of the experiment that was utilized to obtain the DEER spectrum, c) Spin echo data for y-projection (blue), x-projection (orange), and the full length of the state vector (green); the red dot is the optimal place for DEER, d) Data with MW2 turned off (blue), data with MW1 turned off (green), data with both MW1 and MW2 turned on; each data set was calculated by formula (8), x-axis is for MW2 frequency.*

### B. Kinetics of DEER decay

Shifting the pump pulse away from the center reduces the decoherence effect induced on spins A by spins B. In cases when this pulse is in the same place as one of the $\pi/2$ echo pulses, the field caused by B spins becomes the same in both free precession sections, and the DEER effect disappears. By scanning the start time of the spin B pulse with a fixed frequency $f_B$, the kinetics of the DEER signal are revealed (Figure 3a). In Figure 3b, the result of such a 3-pulse DEER sequence is depicted. The plots on the left side are associated with $T < \tau$ and on the right side – with $T > \tau$.

As shown in Figure 3b, besides decay, a significant jump takes place at the point where two $\pi$-pulses overlap. This effect was mentioned in [23], and several ways were suggested to avoid it. We noticed that this distortion appeared in both axis time traces, and also in the angle of the state vector time trace (Figure 3c), but there was no jump in the length of the state vector (Figure 3d). This time trace can be fitted by an exponential decay function to determine the concentration of B spins. In the case of $S_A = S_B = 1/2$, the fitting function is:

$$D = exp(-Cp_B kT) \; ; k = \frac{8\pi^2 \mu_B^2 g^2}{9\sqrt{3}\hbar} \qquad (10)$$

where $C$ – the concentration of B spins, $p_B$ – the probability of spin flip. The 1/2 spin model could be adopted here for spin-1, providing that only 2 sublevels of the NV center with $\Delta m_S = 1$ are utilized.

The existing spin B polarization led to the rotation of the state vector. In [24], it was shown that in this case

$$D = exp(-Cp_B kT) exp(i\varepsilon q \alpha Cp_B kT) \qquad (11)$$

where $\varepsilon$ is the polarization of B spins, $\alpha \approx 0.13213$, and $q$ is a factor that strongly depends on the macroscopic shape of the sample, in our case, on the shape of the part of the diamond illuminated by the laser. Here, the real part of $D$ is associated with the in-phase signal or $D_y$, and the imaginary part is associated with the out-of-phase signal or $D_x$. The value of the slope $Cp_B k = 6.3 \cdot 10^3 \, s^{-1}$ was obtained from the experimental data fit of data in Figure 3d.

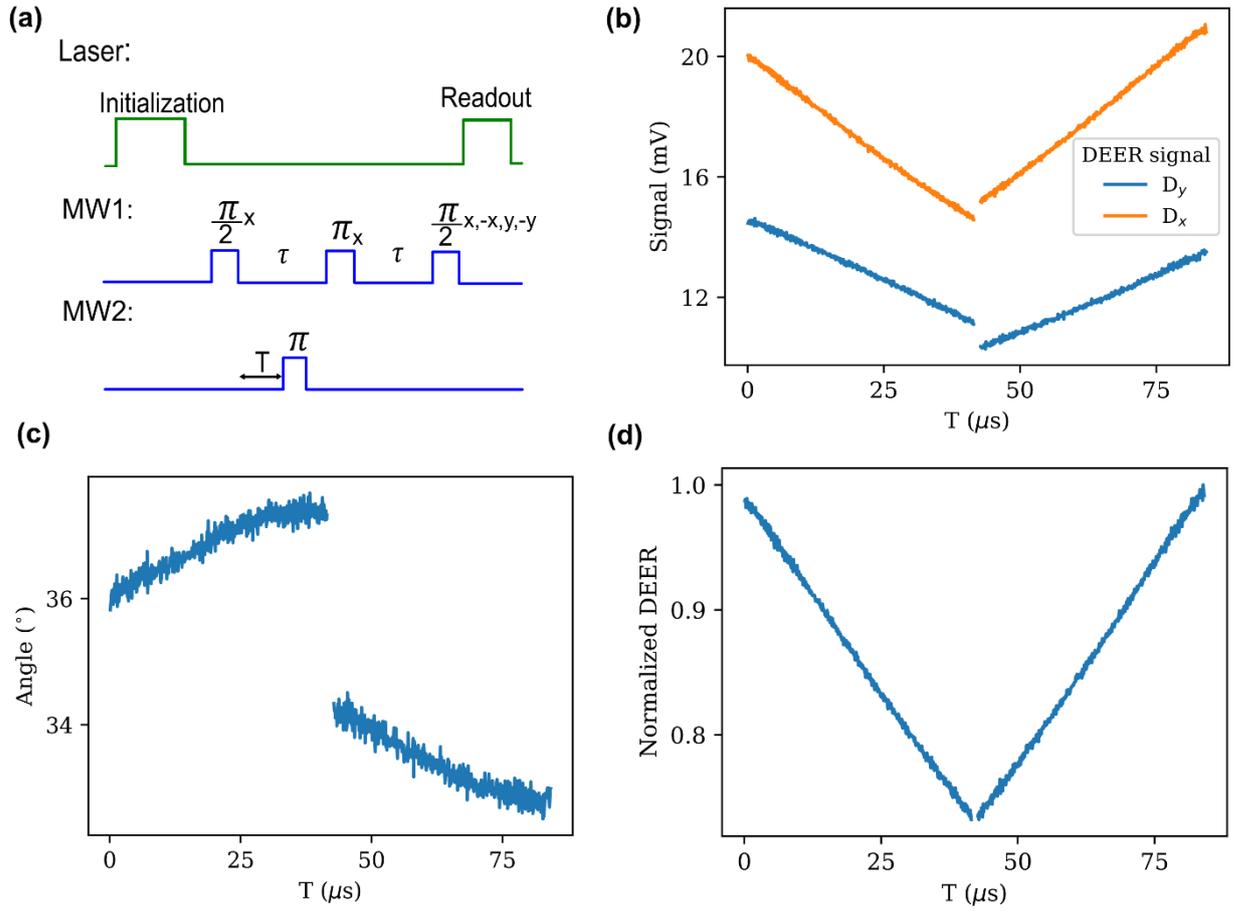

*Figure 3. a) Scheme of a 3-pulse DEER sequence, b) DEER decay time traces calculated using formulas (8) and (9), c) Time trace of the state vector angle calculated using formula (11), d) Time trace of the normalized state vector length calculated using formula (10).*

To describe 3-pulse data distortion, let us consider how the pulse of spin B changes the state of spin A. On the left side of the plot, one can see the result of rotation followed by reflection, on the right side, vice versa, reflection followed by rotation. Neglecting decoherence, during the experiment, the state vector rotates along the equator of the Bloch sphere:

$$|\psi\rangle = \begin{pmatrix} 1 \\ e^{i\varphi} \end{pmatrix} \qquad (12)$$

matrix of $\pi$-pulse:

$$R_\pi = \begin{pmatrix} 0 & -i \\ -i & 0 \end{pmatrix} \qquad (13)$$

matrix of detuned pulse:

$$R_\delta = \begin{pmatrix} e^{\frac{i\delta t}{2}}\left(\cos\left(\frac{\Omega_A t}{2}\right) - i\frac{\delta}{\Omega_A}\right) & -i\frac{\Omega}{\Omega_A}\sin\left(\frac{\Omega_A t}{2}\right)e^{\frac{i\delta t}{2}} \\ -i\frac{\Omega}{\Omega_A}\sin\left(\frac{\Omega_A t}{2}\right)e^{-\frac{i\delta t}{2}} & e^{-\frac{i\delta t}{2}}\left(\cos\left(\frac{\Omega_A t}{2}\right) + i\frac{\delta}{\Omega_A}\sin\left(\frac{\Omega_A t}{2}\right)\right) \end{pmatrix}. \quad (14)$$

$\Omega$ – Rabi frequency, $\delta$ – detuning, $\Omega_A = \sqrt{\Omega^2 + \delta^2}$ – generalized Rabi frequency, $t$ – pulse duration. In approximation $\Omega/\delta \ll 1$, this matrix can be written as:

$$R_\delta = \begin{pmatrix} e^{\frac{i(\delta-\Omega_A)t}{2}} & 0 \\ 0 & e^{\frac{i(\delta-\Omega_A)t}{2}} \end{pmatrix} \quad (15)$$

In the first case, the operator $R_\pi$ follows $R_\delta$; the final state is:

$$|\psi_1\rangle = R_\pi R_\delta |\psi\rangle = \begin{pmatrix} e^{i\left(\varphi - \frac{(\delta-\Omega_A)t}{2} - \frac{\pi}{2}\right)} \\ e^{i\left(\frac{(\delta-\Omega_A)t}{2} - \frac{\pi}{2}\right)} \end{pmatrix} \quad (16)$$

In the second case, vice versa, the operator $R_\delta$ follows $R_\pi$; the final state is

$$|\psi_2\rangle = R_\delta R_\pi |\psi\rangle = \begin{pmatrix} e^{i\left(\varphi + \frac{(\delta-\Omega_A)t}{2} - \frac{\pi}{2}\right)} \\ e^{i\left(-\frac{(\delta-\Omega_A)t}{2} - \frac{\pi}{2}\right)} \end{pmatrix} \quad (17)$$

The phase difference between the states $|\psi_1\rangle$ and $|\psi_2\rangle$ is

$$\Delta\varphi = 2t(\Omega_A - \delta) \approx \Omega^2 t \delta \quad (18)$$

For the values $t = 0.4$ μs, $\delta = 2\pi \cdot 49$ MHz, $\Omega = 2\pi \cdot 1.5$ MHz, the angle shift is $\Delta\varphi = 6°$, while the experimental value is $\Delta\varphi = 4°$. This discrepancy could be explained by the fact that the Rabi frequency is defined from the oscillations of spins B, while it is necessary to put the Rabi frequency obtained at the frequency $f_B$ for spins A. The difference in Rabi frequencies is due to the different orientations of spins A and spins A in relation to the MW antenna magnetic field.

Avoiding a phase jump is possible by using the 4-pulse DEER scheme (see Figure 4a). The sequence contains one more refocusing π pulse, with the pump pulse scanning between these two π pulses. The τ values were chosen to match those used in the 3-pulse scheme. Due to the refocusing pulse, the contrast of such a time trace has an opposite sign. Also, a longer duration of the experiment increases the decoherence effect on the DEER data. This scheme does not contain pulse overlap, and, consequently, there is a phase jump in the DEER signal. To obtain normalized data for the DEER decay in the 3-pulse sequence, one should divide all data by $D(T=0)$, but at this point, the pump pulse is located close to the $\pi/2$-pulse, so their interaction affects the final result. In 4-pulse DEER, one should divide by $D(T=\tau)$, and this value is not distorted by other pulses, so the error of such data is less than in the 3-pulse scheme (see Figure 4). It is worth noting that 4-pulse and 3-pulse sequences give almost the same decay rates when the same experimental parameters are used; thus, the phase jump does not really affect the remaining measurements.

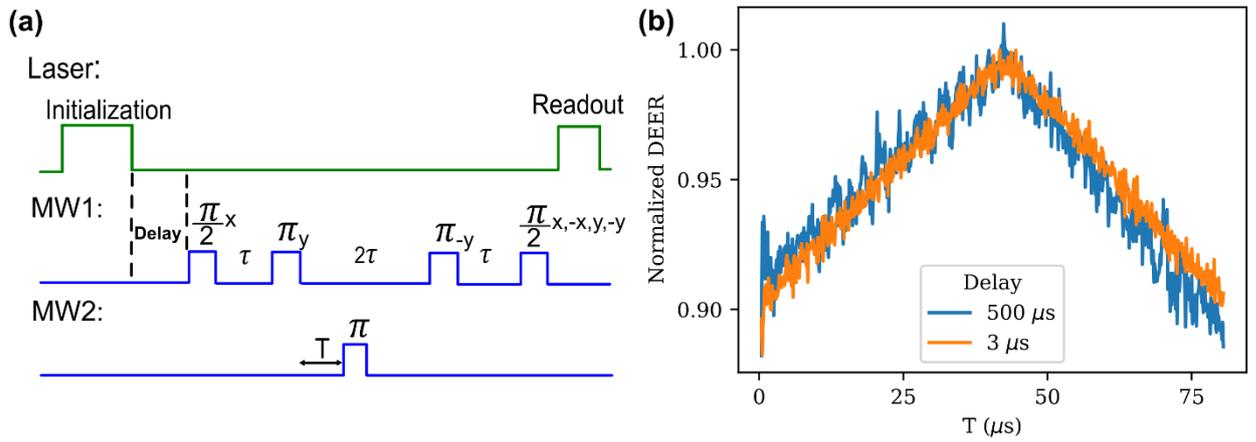

Figure 4. a) Scheme of the 4-pulse DEER experiment, b) 4 pulse DEER time traces with different initial electron polarization levels, both calculated by formula (10).

According to expression (11), the state vector decay rate does not depend on spin polarization and consequently does not depend on focusing and laser power. To verify this, experiments with different electron spin polarizations were conducted (see Figure 5). After a laser polarization pulse, electron spin state thermalization with $T_1$ begins. To change the initial polarization, the delay between the laser pulse and the MW sequence was prolonged to 500 µs, reducing polarization by 35%, as it is for operational 3 µs. Essentially, the signal-to-noise ratio degraded, but the decay rate changed within fit parameter error.

There are two possible approaches to experimentally determine the coefficient $k$ from formula (10) for NV-NV DEER. The first approach involves performing the described experiments on different diamond samples with known NV concentrations, previously measured via optical spectroscopy, and fitting the decay rate as a function of concentration using the same $C\, p_B$. The second approach is to scan $p_B$, that calculated with formula (2) and to fit the decay rate from the concentration multiplied by the $p_B$ dependence on the sample with a known NV concentration. According to the existing theory of DEER decay kinetics, both methods should yield the same result; however, the second approach was chosen due to its practical simplicity The results of this experiment, conducted on a diamond sample with an NV concentration of 0.7 ppm, are shown in Figure 5. The decay rate dependence on the excited concentration was fitted using a linear function $y = kx$ with $k = 34\ \text{s}^{-1}\text{ppb}^{-1} = 1.94\ 10^{-13}\ \text{s}^{-1}\text{cm}^{-3}$. This value is 8 times smaller than the prediction from formula (10). This discrepancy may be attributed to the $S=1$ feature or NV-NV interaction that may be not fully described by the dipole formula; however, the exact cause remains unknown.

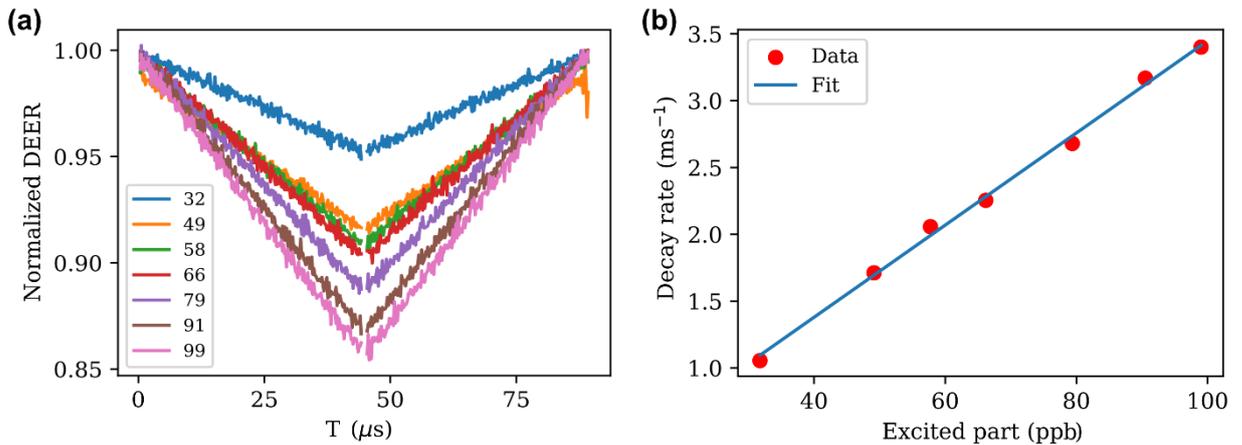

*Figure 5. a) Data of DEER time traces where different fractions of B spins were excited, the value in the legend corresponds to the concentration of excited B spins in ppb units, b) dependence of the DEER decay rate on the concentration of excited B spins.*

To study the evolution of spins A in the DEER experiment in more detail, we also calculated the angle jump using formula (9) and compared the experimental results with predictions from (11) and (12). In the works [24,25], a similar evolution was observed at a cryogenic temperature and a high magnetic field. In the case of NV centers, it is possible to reach high polarization at a low magnetic field and room temperature by laser illumination. The rotation dynamics of the state vector can be understood from Figure 6a, where the fraction of excited NV centers in group B was changed. According to formula (12), the angle distortion should change with Rabi frequency in a linear manner, if the duration of the pulse corresponds to the $\pi$-pulse $t = \pi/\Omega$. The experimental data fit gave us a slope of 4.4 °/MHz (see Figure 6b), while formula (12) gave a value of 7.1 °

/MHz. The angular velocity also depends on the excitation level and decay rate; consequently, formula (11) predicts a slope of $0.13\,\varepsilon$ in the case of a spherical sample shape, while the experimental value is 0.43 Although NV polarization under laser illumination may reach values as high as 0.9, accurately estimating the polarization is challenging due to the unknown radial intensity distribution of the laser beam. This uncertainty makes it difficult to determine the exact shape of the illuminated volume within the diamond. However, certain sample geometries and orientations relative to the magnetic field may result in such high polarization values, as suggested in [25].

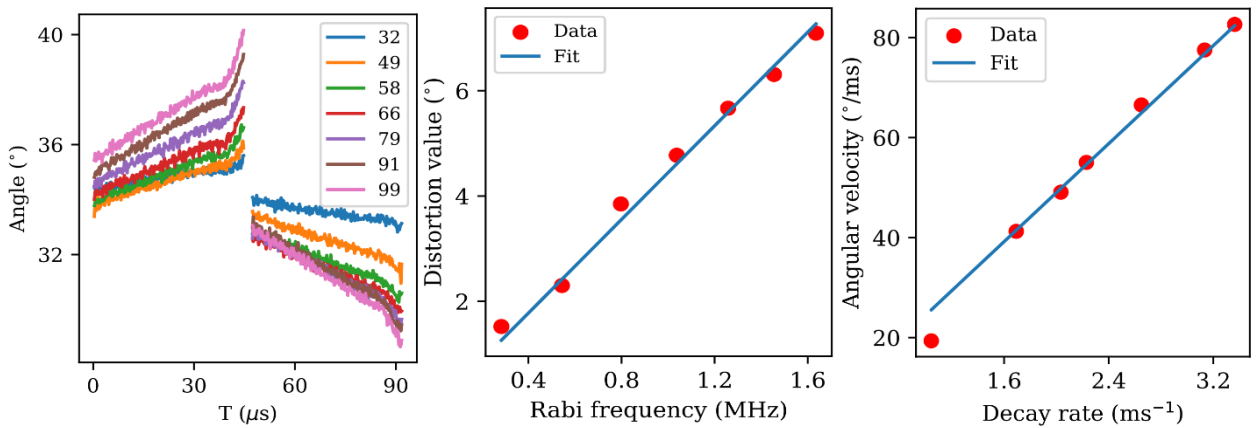

*Figure 6. a) Angle time traces with different excitation levels, the value in the legend corresponds to the concentration of excited B spins in ppb units, b) Distortion value from Rabi frequency dependence, c) Angular velocity of the state vector from the DEER decay rate dependence.*

## CONCLUSION AND OUTLOOK

In this work, a direct measurement of the mutual interaction between NV centers in a dense ensemble was experimentally demonstrated using a dynamical DEER sequence. Two types of DEER sequences were investigated: a 3-pulse and a 4-pulse scheme. The phase jump observed in the 3-pulse DEER sequence was analytically shown to result from non-commuting operations within the pulse sequence. This explanation is supported by the experimental dependence of the phase jump on microwave power. The observed angular dynamics of the spin state vector are consistent with the existing theory of out-of-phase DEER. The 4-pulse DEER scheme, in contrast, is not affected by phase jumps, as theoretically expected—and this was confirmed experimentally. The interaction dynamics between NV centers were studied as a function of effective NV concentration, controlled by varying the number of flipped "bath" spins in the dynamical DEER sequence.

## IV.  ACKNOWLEDGMENTS

We thank Professor Alexander P. Nizovzev for fruitful discussions. This study was supported by the Ministry of Science and Higher Education of the Russian Federation (agreement/grant No. 075-15-2024-556).

## V.  REFERENCES

[1] E.L. Hahn, Spin Echoes, Physical Review 80 (1950) 580–594. https://doi.org/10.1103/PhysRev.80.580.

[2] A.D. Milov, K.M. Salikhov, M.D. Shirov, Application of ELDOR in electron-spin echo for paramagnetic center space distribution in solids, Fizika Tverdogo Tela 23 (1981) 975–982.

[3] A.D. Milov, A.B. Ponomarev, Y.D. Tsvetkov, Electron-electron double resonance in electron spin echo: Model biradical systems and the sensitized photolysis of decalin, Chem Phys Lett 110 (1984) 67–72. https://doi.org/10.1016/0009-2614(84)80148-7.

[4] R.E. Martin, M. Pannier, F. Diederich, V. Gramlich, M. Hubrich, H.W. Spiess, Determination of End-to-End Distances in a Series of TEMPO Diradicals of up to 2.8 nm Length with a New Four-Pulse Double Electron Electron Resonance Experiment, Angewandte Chemie International Edition 37 (1998) 2833–2837. https://doi.org/https://doi.org/10.1002/(SICI)1521-3773(19981102)37:20<2833::AID-ANIE2833>3.0.CO;2-7.

[5] B. Joseph, E.A. Jaumann, A. Sikora, K. Barth, T.F. Prisner, D.S. Cafiso, In situ observation of conformational dynamics and protein ligand–substrate interactions in outer-membrane proteins with DEER/PELDOR spectroscopy, Nat Protoc 14 (2019) 2344–2369. https://doi.org/10.1038/S41596-019-0182-2;SUBJMETA.

[6] T. Bahrenberg, Y. Rosenski, R. Carmieli, K. Zibzener, M. Qi, V. Frydman, A. Godt, D. Goldfarb, A. Feintuch, Improved sensitivity for W-band Gd(III)-Gd(III) and nitroxide-nitroxide DEER measurements with shaped pulses, Journal of Magnetic Resonance 283 (2017) 1–13. https://doi.org/10.1016/J.JMR.2017.08.003.

[7]     R.G. Larsen, D.J. Singel, Double electron–electron resonance spin–echo modulation: Spectroscopic measurement of electron spin pair separations in orientationally disordered solids, J Chem Phys 98 (1993) 5134–5146. https://doi.org/10.1063/1.464916.

[8]     S.H. White, ed., Membrane Protein Structure, Springer New York, New York, NY, 1994. https://doi.org/10.1007/978-1-4614-7515-6.

[9]     D.B. Gophane, S.T. Sigurdsson, TEMPO-derived spin labels linked to the nucleobases adenine and cytosine for probing local structural perturbations in DNA by EPR spectroscopy, Beilstein Journal of Organic Chemistry 11:24 11 (2015) 219–227. https://doi.org/10.3762/BJOC.11.24.

[10]    O. Schiemann, N. Piton, Y. Mu, G. Stock, J.W. Engels, T.F. Prisner, A PELDOR-Based Nanometer Distance Ruler for Oligonucleotides, J Am Chem Soc 126 (2004) 5722–5729. https://doi.org/10.1021/JA0393877.

[11]    M.W. Doherty, N.B. Manson, P. Delaney, F. Jelezko, J. Wrachtrup, L.C.L. Hollenberg, The nitrogen-vacancy colour centre in diamond, Phys Rep 528 (2013) 1–45. https://doi.org/10.1016/j.physrep.2013.02.001.

[12]    F.H. Cho, V. Stepanov, C. Abeywardana, S. Takahashi, 230/115 GHz Electron Paramagnetic Resonance/Double Electron–Electron Resonance Spectroscopy, Methods Enzymol 563 (2015) 95–118. https://doi.org/10.1016/BS.MIE.2015.07.001.

[13]    V. Stepanov, S. Takahashi, Determination of nitrogen spin concentration in diamond using double electron-electron resonance, Phys Rev B 94 (2016) 024421. https://doi.org/10.1103/PhysRevB.94.024421.

[14]    O.R. Rubinas, V.V. Soshenko, S.V. Bolshedvorskii, I.S. Cojocaru, A.I. Zeleneev, V.V. Vorobyov, V.N. Sorokin, V.G. Vins, A.N. Smolyaninov, A.V. Akimov, Optical detection of an ensemble of C centres in diamond and their coherent control by an ensemble of NV centres, Quantum Elec (Woodbury) 51 (2021) 938–946. https://doi.org/10.1070/QEL17624/XML.

[15]    O.R. Rubinas, V. V. Soshenko, S. V. Bolshedvorskii, I.S. Cojocaru, V. V. Vorobyov, V.N. Sorokin, V.G. Vins, A.P. Yeliseev, A.N. Smolyaninov, A. V. Akimov, Optimization of the Double Electron–Electron Resonance for C-Centers in Diamond, Physica Status Solidi (RRL) – Rapid Research Letters 16 (2022) 2100561. https://doi.org/10.1002/PSSR.202100561.

[16] S. Li, H. Zheng, Z. Peng, M. Kamiya, T. Niki, V. Stepanov, A. Jarmola, Y. Shimizu, S. Takahashi, A. Wickenbrock, D. Budker, Determination of local defect density in diamond by double electron-electron resonance, Phys Rev B 104 (2021) 094307. https://doi.org/10.1103/PHYSREVB.104.094307/FIGURES/8/MEDIUM.

[17] T. Joas, F. Ferlemann, R. Sailer, P.J. Vetter, J. Zhang, R.S. Said, T. Teraji, S. Onoda, T. Calarco, G. Genov, M.M. Müller, F. Jelezko, High-Fidelity Electron Spin Gates for Scaling Diamond Quantum Registers, Phys Rev X 15 (2025) 021069. https://doi.org/10.1103/PhysRevX.15.021069.

[18] V. V Soshenko, O.R. Rubinas, V. V Vorobyov, S. V Bolshedvorskii, P. V Kapitanova, V.N. Sorokin, A. V Akimov, Microwave Antenna for Exciting Optically Detected Magnetic Resonance in Diamond NV Centers, Bulletin of the Lebedev Physics Institute 45 (2018) 237–240. https://doi.org/10.3103/S1068335618080043.

[19] J.M. Schloss, J.F. Barry, M.J. Turner, R.L. Walsworth, Simultaneous Broadband Vector Magnetometry Using Solid-State Spins, Phys Rev Appl 10 (2018) 034044. https://doi.org/10.1103/PHYSREVAPPLIED.10.034044/FIGURES/5/MEDIUM.

[20] P. Kehayias, M. Mr??zek, V.M. Acosta, A. Jarmola, D.S. Rudnicki, R. Folman, W. Gawlik, D. Budker, Microwave saturation spectroscopy of nitrogen-vacancy ensembles in diamond, Phys Rev B Condens Matter Mater Phys 89 (2014) 1–8. https://doi.org/10.1103/PhysRevB.89.245202.

[21] A. Schweiger, G. Jeschke, Principles of Pulse Electron Paramagnetic Resonance, Oxford University PressOxford, 2001. https://doi.org/10.1093/oso/9780198506348.001.0001.

[22] P.L. Stanwix, L.M. Pham, J.R. Maze, D. Le Sage, T.K. Yeung, P. Cappellaro, P.R. Hemmer, A. Yacoby, M.D. Lukin, R.L. Walsworth, Coherence of nitrogen-vacancy electronic spin ensembles in diamond, Phys Rev B 82 (2010) 201201. https://doi.org/10.1103/PhysRevB.82.201201.

[23] S.A. Dzuba, Conducting a three-pulse DEER experiment without dead time: A review, J Magn Reson Open 14–15 (2023) 100100. https://doi.org/10.1016/J.JMRO.2023.100100.

[24] S.R. Sweger, V.P. Denysenkov, L. Maibaum, T.F. Prisner, S. Stoll, The effect of spin polarization on double electron-electron resonance (DEER) spectroscopy, Magnetic Resonance 3 (2022) 101–110. https://doi.org/10.5194/MR-3-101-2022.

[25] V. Denysenkov, T.F. Prisner, P. Neugebauer, S. Stoll, A. Marko, Macroscopic sample shape effect on pulse electron double resonance (PELDOR) signal, Journal of Magnetic Resonance 356 (2023) 107564. https://doi.org/10.1016/J.JMR.2023.107564.